\documentclass[journal=jacsat,manuscript=article]{achemso}

\setkeys{acs}{keywords=true}
\usepackage{chemformula} 
\usepackage[T1]{fontenc} 

\author{Nanyang Xu}
\email{nyxu@zhejianglab.edu.cn}
\affiliation{Research Center for Quantum Sensing, Zhejiang Lab, Hangzhou, 311000, China}
\alsoaffiliation{School of Physics,Hefei University of Technology, Hefei, Anhui 230009, China}

\author{Feifei Zhou}
\affiliation{Research Center for Quantum Sensing, Zhejiang Lab, Hangzhou, 311000, China}
\alsoaffiliation{School of Physics,Hefei University of Technology, Hefei, Anhui 230009, China}

\author{Xiangyu Ye}
\affiliation{CAS Key Laboratory of Microscale Magnetic Resonance, University of Science and Technology of China, Hefei 230026, China}

\author{Xue Lin}
\affiliation{Research Center for Quantum Sensing, Zhejiang Lab, Hangzhou, 311000, China}
\alsoaffiliation{School of Physics,Hefei University of Technology, Hefei, Anhui 230009, China}

\author{Bao Chen}
\affiliation{Research Center for Quantum Sensing, Zhejiang Lab, Hangzhou, 311000, China}
\alsoaffiliation{School of Physics,Hefei University of Technology, Hefei, Anhui 230009, China}

\author{Ting Zhang}
\affiliation{School of Physics,Hefei University of Technology, Hefei, Anhui 230009, China}

\author{Feng Yue}
\affiliation{Engineering Research Center of Safety Critical Industrial Measurement and Control Technology, Ministry of Education, Hefei  230009}

\author{Bing Chen}
\affiliation{School of Physics,Hefei University of Technology, Hefei, Anhui 230009, China}

\author{Ya Wang}
\email{ywustc@ustc.edu.cn}
\affiliation{CAS Key Laboratory of Microscale Magnetic Resonance, University of Science and Technology of China, Hefei 230026, China}

\author{Jiangfeng Du}
\email{djf@ustc.edu.cn}
\affiliation{CAS Key Laboratory of Microscale Magnetic Resonance, University of Science and Technology of China, Hefei 230026, China}

\title{Noise prediction and reduction of single electron spin by deep-learning-enhanced feedforward control}

\abbreviations{IR,NMR,UV}

\begin{document}


\begin{abstract}
Noise-induced control imperfection is an important problem in applications of diamond-based nano-scale sensing, where measurement-based strategies are generally utilized to correct low-frequency noises in realtime. However, the spin-state readout requires a long time due to the low photon-detection efficiency. This inevitably introduces a delay in noise-reduction process and limits its performance. Here we introduce the deep learning approach to relax this restriction by predicting the trend of noise and compensating the delay. We experimentally implement feedforward quantum control of nitrogen-vacancy center in diamond to protect its spin coherence and improve the sensing performance against noise. The new approach effectively enhances the decoherence time of the electron spin, which enables exploring more physics from its resonant spectroscopy. A theoretical model is provided to explain the improvement. This scheme could be applied in general sensing schemes and extended to other quantum systems.
\end{abstract}

Noise control and suppression lies at the heart of developing practical quantum technologies. Many methods are proposed and implemented in various applications, including open-loop techniques such as optimal control \cite{nv_qec,quant_entangle_2014} and dynamical decoupling (DD)\cite{du_nature_2009,hanson_nv_DD_2010, dd_2011_psi}, as well as closed-loop approaches like the feedback control. In spite of the great advances, these methods usually require precise control conditions and limit their potential applications. For example, the recently emergent nano-scale magnetic resonance imaging technique based on nitrogen-vacancy (NV) centers in diamond will have higher magnetic sensitivity in higher magnetic fields \cite{imaging2012Walsworth, imaging2013Lukin, superconducting2016Maletinsky,imaging2019jd},but the instability of the magnets severely limits the sensitivity improvement due to the resulting significant resonance frequency fluctuation for a strong magnetic field on the order of Tesla. A method that can realtime learn these external noises, update the control, and naturally combine with the well-developed control ways will thus enhance the power of quantum technologies towards practical applications.

As a popular topic in scientific researches, machine learning (ML) is a group of methods that are well developed in computer science and widely applied in many other areas. For example in the quantum physics fields, ML method has been used for Hamiltonian learning \cite{Hlearning_nphys2017}, qubit readout \cite{readout_ML_superconduct,readout_ML_ion, ml_readout_2021apl, ml_readout_2010,Fabio_prapp_2018}, noise reduction \cite{nv_ml_sr_2019} and experimental controls \cite{auto_exp_PRL2016,ML_predict_new_exp}. Recently, the deep learning model (DL) \cite{deep_learning_science} is emerging as a powerful tool, which can deal with complicated problems with multiple layers of artificial neural network such as the many-body problem \cite{science_2017_manybody}. Another potential application is to combine DL with quantum feedforward control by using its ability in time-series prediction \cite{timeseries_predict,time_forecast_2013_scibul}, which has been applied in stock-price prediction and weather forecast.

Here we propose a DL-assisted feedforward control scheme and show its advantage on a solid-state single spin associated with the nitrogen-vacancy (NV) center in diamond. In our scheme, the target spin works as a quantum sensor to detect the magnetic noise used to train the DL agent. The agent learns the noise model and predicts the forthcoming value in realtime and feeds this value to correct the corresponding error on the spin. Due to the low optical detection efficiency, there's an intrinsic latency in magnetic-noise readout process in the order of seconds. By using the new scheme, we demonstrate a feedforward Ramsey experiment and show a three-fold (50\%) improvement in the linewidth of its spectrum as well as the spin decoherence time comparing with conventional (non-DL feedback) scheme. As an application of this technique, we resolve a group of resonant lines in Ramsey spectroscopy which are indistinguishable in the conventional scheme under a magnetic field of 500 G. This scheme can be easily extended to ensemble NV centers and other quantum control systems in the future.


\begin{figure*}[htb]
\centering
\includegraphics[width=0.95\textwidth]{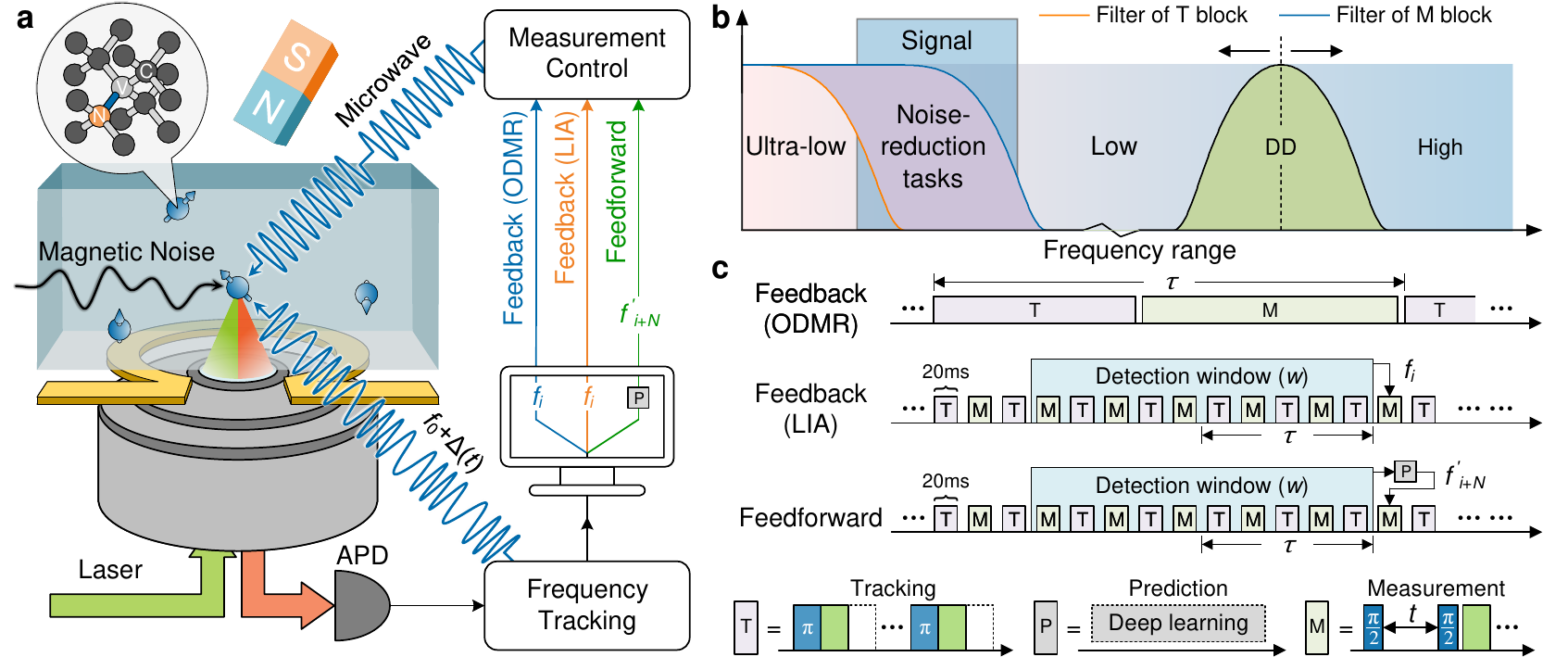}
\caption{\textbf{Idea of feedforward quantum control.} \textbf{a)} Schematic of the control scheme and experimental setup. A frequency tracking process tracks the resonance frequency $f_i$ of electron spin using LIA or ODMR method at each time step $i$, where the microwave is swept in the adjacent range $\Delta$ of an initial frequency $f_0$ and the emitting fluorescence is collected. A control computer reads the detected value and feeds the detected (predicted) frequency to the measurement controller for the subsequent experimental loops. We realize three kinds of experimental schemes: feedback with ODMR (blue arrow), feedback with LIA (orange arrow), and feedforward (green arrow), respectively. \textbf{b)} Noise reduction in the frequency domain. The orange (blue) curve exhibits the noise (noise + signal) obtained via noise tracking (measurement) like a low-pass filter, while DD-based sensing protocol works as a band-pass filter. \textbf{c)} Detailed control sequences of the experimental schemes. The notation \textbf{T} represents pulse sequence of frequency detection and \textbf{M} is general-purpose measurement, where Ramsey measurement is used as an example and $t$ is the free evolution time. $\tau$ is defined as the period of frequency tracking in ODMR or the detection delay in LIA. $w$ is the time length of detection window in LIA. Label \textbf{P} denotes DL agent that predicts the future values $f_{i+N}^{'}$ based on history data, where $N$ is the prediction length.}
\label{fig:idea}
\end{figure*}

NV center in diamond has been developed as a multi-function sensor for magnetic field \cite{Maze2008, Balasubramanian2008, high_d_range, Grinolds2013, 1602.07144, magnetometer2016jd, Joerg_broadband,magnetometer2021jd}, electric field \cite{nv_electric_sensing,lirui_prl_2020} and temperature \cite{Kucsko2013,Toyli2013}. In particular, the development of the dynamical decoupling (DD) control method \cite{DD_opensys_prl_2019}, which acts as a band-pass noise filter working at medium frequency (Fig.~\ref{fig:idea}b), advances NV centers as an outstanding sensors for nano-scale magnetic resonances and imaging (nMRI) \cite{fedor_science_2017, Joerg_NMR2013, Rugar_NMR_2013, fazhan_nmehtods2018, Joerg_chemical_shift}. On the other hand, due to the high gyromagnetic ratio, NV centers suffers from low-frequency magnetic noise arising from environmental factors, for example, the magnetic noise due to mechanical vibrations or temperature changes. The low-frequency magnetic noise also imposes a limitation for improving the sensitivity of nMRI by directly enhancing the magnetic fields like conventional nuclear magnetic resonance. A noise-suppression magnetic detection working at the low frequency regime will thus benefit those applications. Our DL-assisted feedforward control scheme provides a novel way to address these limitation.

The control architecture and principle of a measurement-based NV experiment is shown in Fig.~\ref{fig:idea}a-b. In a DC sensing where the useful information (signal) often lays in the low-frequency domain. All the information is obtained from the measurement (M) and the high-frequency noise is  filtered out by signal accumulations. Meanwhile, a feedback process can be realized to compensate the ultra-low-frequency noises ( \emph{e.g.}, the $1/f$ or $1/f^2$ noise ) by tracking (T) the electron-spin resonant frequency independently of the measurement. On the other hand, the signal accumulation introduces a latency in the detected noise thus the efficiency of feedback is therefore harmed.

\begin{figure}[tb]
    \centering
    \includegraphics[width=0.55\textwidth]{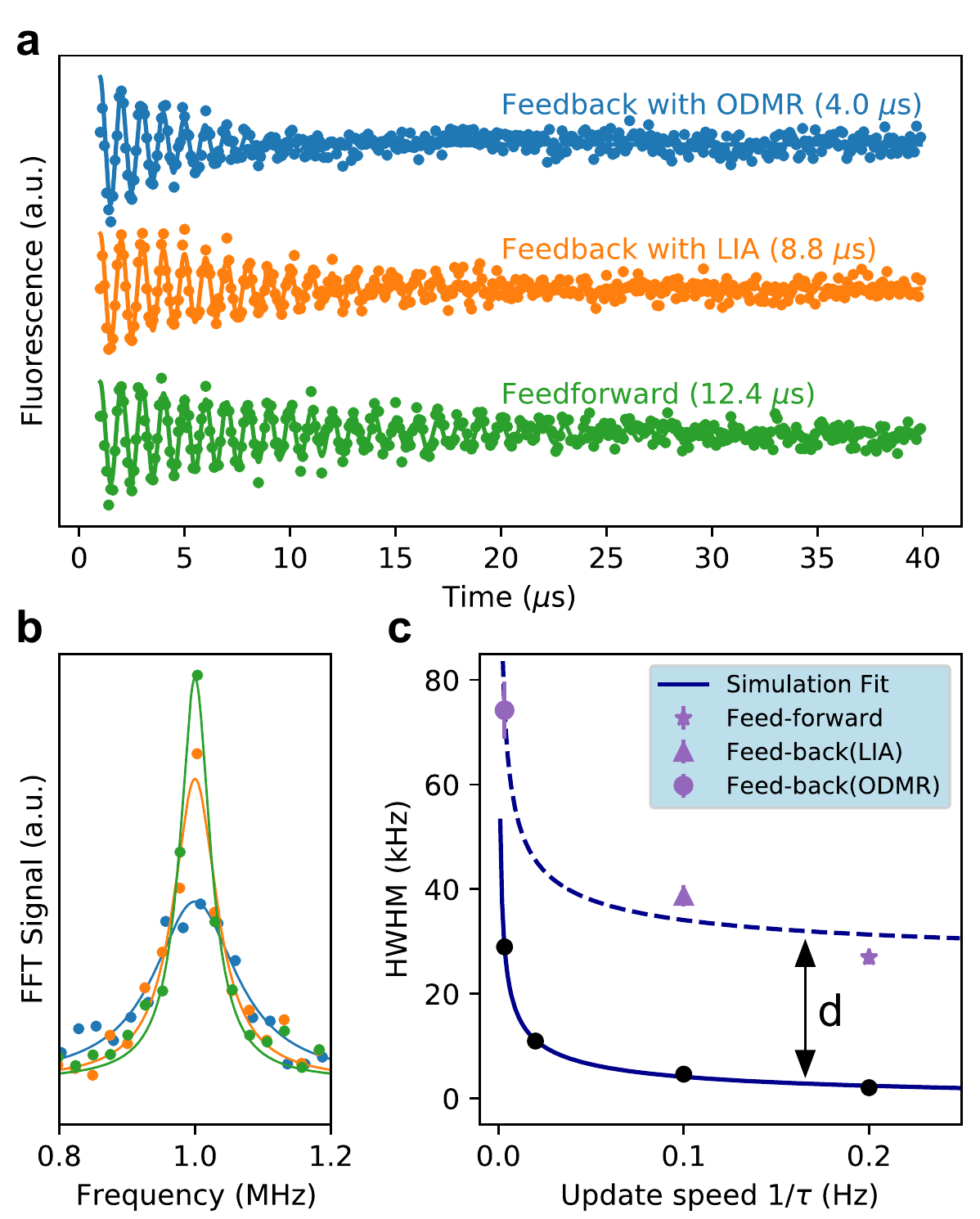}
    \caption{\textbf{Expected performance of the measurement-based noise reduction.} \textbf{a-b)} Experimental results (a) and their FFT spectra (b) of Ramsey measurement on NV1 with a bias of $1$ MHz off the resonance using the three specific experimental schemes in Fig.~\ref{fig:idea}a. The bracketed time in (a) is the fitted spin decoherence time $T_2^{*}$. \textbf{c)} The dependence of Ramsey linewidth (HWHM) $l$ on the update speed $\nu=1/\tau$. The dots (shapes) are fitted with the same function $l=1/\nu^{n}$($l=1/\nu^{n}+d$). All simulations are based on the recorded noise in Fig.~\ref{fig:analysis}a.}
    \label{fig:main_result}
\end{figure}

We consider three measurement-based schemes and use Ramsey measurement as an example. The experiment  is performed on a single NV center using a home-built confocal microscope system (see Supporting Information). Each control scheme has a different detection latency (\emph{i.e.}, the update period $\tau$ defined in Fig.~\ref{fig:idea}) thus associates with different noise-reduction efficiencies. The first one is feedback control using conventional optically-detected magnetic resonance (ODMR) method without prediction, where the resonant frequency is detected by sweeping out an ODMR spectrum with a range of microwave (MW) frequencies and finding the spectral peak. Although this scheme is commonly used in NV-based experiments, it is quite slow with a latency time often over hundreds of seconds and the result is shown in upper plot of Fig.~\ref{fig:main_result}a. With some advanced protocols\cite{dual_iso_B_1, dual_iso_B_2, dual_iso_B_3, dual_iso_B_4}, however, sweeping the whole line shape is not necessary to detect resonances. In the second scheme, we adopt the lock-in amplifier (LIA ) method\cite{lock_in_1, lock_in_2} to improve the efficiency of frequency detection. Unlike ODMR, LIA gets the resonate frequency directly by a hardware synchronization of photon counts and MW frequencies via an external oscillating reference. It scans only a small range near the resonant frequency, thus enables us updating the detection in realtime and tracking the magnetic noise continuously. This scheme also uses the feedback strategy and the result is shown in the middle plot of Fig.~\ref{fig:main_result}a, which has a two-fold improvement in the spin decoherence time $T_2^{*}$ comparing with the first scheme.

\begin{figure}[htb]
\centering
\includegraphics[width=0.6\textwidth]{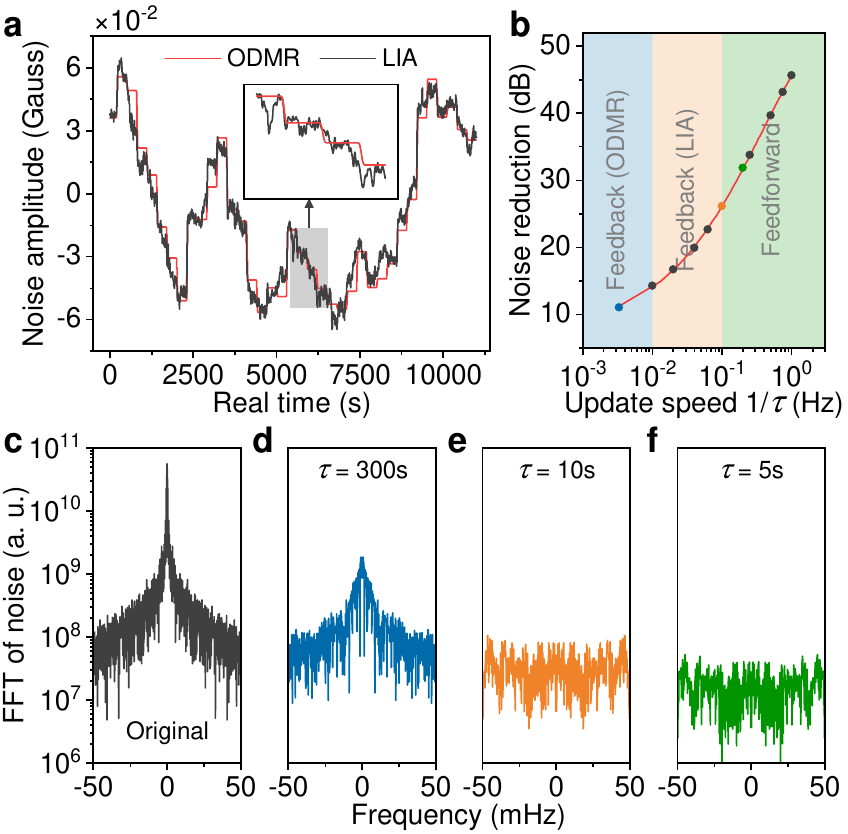}
\caption{ \textbf{Analysis of magnetic noise reduction.} \textbf{a)} A free-run tracking of magnetic noise on NV1. The continuous line (black) is recorded using LIA, and is interrupted by ODMR (red square) for every 5 minutes. The update period $\tau$ of ODMR (LIA) method is fixed at 300 (1) s specifically. \textbf{b)} Noise reduction efficiency in an ideal feedback control (with no detection delay). The filled areas in different colors represent the update speeds that can be realized in the specific control schemes, \emph{i.e.}, ODMR-based (blue) and LIA-based feedback (orange), as well as feedforward (green). \textbf{c-f)} FFT spectra of the noise before (c) and after (d-e) feedback or (f) feedforward control, where (d-f) associate with the update speed 0.0033, 0.1, and 0.2 Hz in (b), respectively.}
\label{fig:analysis}
\end{figure}

\begin{figure*}[htb]
\centering
\includegraphics[width=1\textwidth]{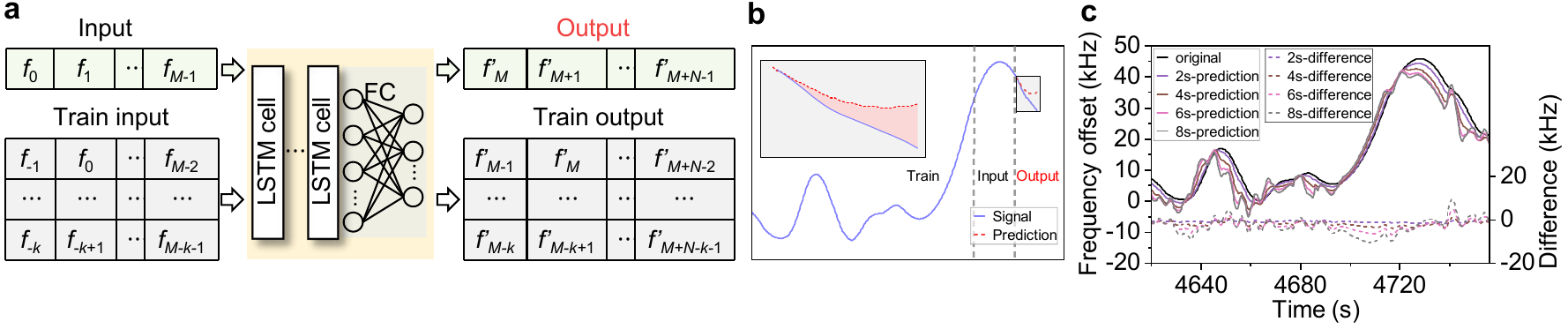}
\caption{\textbf{Noise prediction using deep learning.} \textbf{a)} Architecture of the model. The DL agent consists of basic LSTM cells followed by a  fully-connected (FC) network used to bridge layers with different dimensions. It uses recent $M$ values of $f_t$ as input and generates $N$ predicted values of $f^{'}_t$. The history data for training is divided into $k$ sequences, each of which is shifted forward one value for each step. \textbf{b)} The data diagram in noise prediction. The inset plot highlights the difference between original and predicted signals. \textbf{c)} The comparison and difference (the secondary y axis) between original and predicted noise with different prediction lengths $N$.}
\label{fig:predict_model}
\end{figure*}

Because of low photon counts of single-spin system, LIA use a detection window around 20 s for signal accumulation. This latency fundamentally limits the efficiency of feedback control. In the third scheme, feedforward strategy is utilized to further improve the performance assisted with the DL method. The feedforward strategy, different from feedback control that uses detected value directly, predicts the trend of noise in the future and uses the forthcoming value instead to correct the error. The result is shown in the lower plot of Fig.~\ref{fig:main_result}a, where the performance is about three-fold improved comparing with the first scheme. We also plot the corresponding FFT spectra in Fig.~\ref{fig:main_result}b and the linewidth $l$ (\emph{i.e.}, the half width of the peak at half its maximum, HWHM) is as well improved where $l \approx 1/\pi T_2^{*}$ in our case.

To understand this improvement, first we perform a free-run tracking of the magnetic noise from the experiment for a long time as shown in Fig.~\ref{fig:analysis}a. Then we realize a numerical simulation based on the recorded noise to mimic the feedback (feedforward) process. In the simulation, the amplitude used to correct the instant noise is updated in a period of $\tau$, and the noise-reduction efficiency depends on $\tau$ closely as shown in Fig.~\ref{fig:analysis}b. The corresponding $\tau$ in the above three schemes refers to 300 s, 10 s and 5 s specifically, and the FFT spectra of the noise before and after the feedback (feedforward) process are shown in Fig.~\ref{fig:analysis}c-f. One can find that the shorter the latency is, the better the noise can be corrected (Fig.~\ref{fig:analysis}b), which highlights the role of detection latency in measurement-based quantum control. Note that the signal accumulation time in experimental limits the latency time to half the detection window of LIA (\emph{i.e.}, $\tau$ = $w$/2 =10 s ). And the update period of 5 s in the third scheme is taken into account the prediction of the noise that is discussed in the following. Beside the detection latency, a full consideration of the impacts on the noise reduction efficiency is discussed in detail, as well as the performance on different types of noise (see Supporting Information).

The noise-reduction effect on the Ramsey measurement can also be figured out from the linewidth of the Ramsey spectrum. Before analyzing the experimental results, we simulate the measurements and analyze the broadening of linewidth when the noise in Fig.~\ref{fig:analysis}a is applied (see Supporting Information). For different update speeds $\nu=1/\tau$, the fitted linewidth (HWHM) $l$ is shown in Fig.~\ref{fig:main_result}c, where a relation $l = 1/\nu^{n}$ is used to fit the data (the solid line). The fitted $n=0.48$ represent the dependence of the inhomogeneous broadening on the feedback speed under the given noise. The experimental result matches well with this relation (the dashed line) except for a residual broadening (\emph{i.e.}, $l = 1/\nu^{n} + d$). This broadening $d \sim$ 26 kHz comes from noise due to the high-frequency part outside the frequency range of the LIA detection. Note that this numerical simulation does not include the sensitivity-induced noise when the detection time is very short in a fast feedback process. 

In the feedforward scheme, we adopt the long short-term memory (LSTM) \cite{lstm_1997} (see Supporting Information), \emph{i.e.}, an improved version of the recurrent neural network (RNN) for noise prediction as shown Fig.~\ref{fig:predict_model}a. The work flow of the noise prediction is shown in Fig.~\ref{fig:predict_model}b. In prior, history data recorded from LIA is fed to the agent to train the neural network. In experiment, the agent gets recent detected values of $M$ seconds as input and predicts $N$-second values as output, where $N$ ($M$) is the prediction (sequence) length. We further compare original and predicted signals directly as well as their difference in Fig.~\ref{fig:predict_model}c with different prediction lengths $N$. The difference is shown to become larger and larger with $N$ increasing. This fact indicates a trade-off between the fidelity and prediction length $N$ to compensate the detection delay. For example, in our experiment the best noise reduction performance is achieved when $N$ is about 5 s, while the detection delay is 10 s. Furthermore, benefited from the merit of the LSTM neural network, this prediction fidelity would be better for long-correlated signals ($e.g.$, a periodical disturbance in the noise).

\begin{figure}[tb]
\centering
\includegraphics[width=0.7\textwidth]{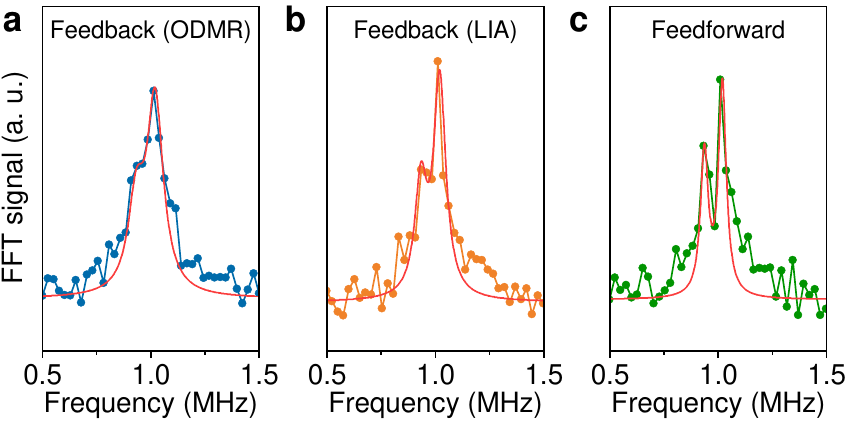}
\caption{\textbf{Application of feedforward Ramsey measurements.} \textbf{a-c)} FFT spectra corresponding to the Ramsey measurement of NV2 using (a) ODMR-based feedback, (b) LIA-based feedback and (c) feedforward schemes.}
\label{fig:nvpair}
\end{figure}

Ramsey measurement has a wide range of applications in magnetic resonance (MR) sciences and sensing. Here as an example, we apply the improved protocol in Ramsey spectroscopy on another NV center (NV2). As shown in Fig.~\ref{fig:nvpair}b-c, according to the fitted results based on Lorentzian function, our method resolves two resonant peaks, which is not visible in conventional experiments even when the ODMR-based feedback strategy is used (Fig.~\ref{fig:nvpair}a). Because the photon counts of NV2 is almost twice of NV1 and no electron-nuclear or electron-electron coupling is observed (see Supporting Information), the most possible case resulting in this spectrum is that NV2 is a pair with two centers aligned in the same direction. The tiny difference in resonant frequency is possibly because of electron-spin zero-field splitting, which cannot be observed by the DD-based techniques. Another possibility is that NV2 is a single center coupled with a parallel nuclear spin along the electron-spin quantization axis.
Under both cases, our method can help observing and selectively driving small-gapped resonance lines, opening the possibility to explore more physics or controllability of NV center in the future.

To conclude, we propose and implement a novel feedforward quantum control scheme by learning the knowledge of low-frequency noise and correct the corresponding error in realtime. The new scheme utilizes a LIA-based approach instead of standard ODMR method because the lock-in detection enables continuous frequency tracking that is required for learning and compensating the noise in realtime. Meanwhile, the LIA detection focuses on the local changes of the resonant frequency, thus the update period $\tau$ is naturally shorter than ODMR, which helps improve the performance. We realize the new scheme as an example in the Ramsey spectroscopy and the linewidth is reduced to 1/3 of  the original, where the magnetic noise mainly comes from the instability of external magnetic field induced by the temperature fluctuation (see Supporting Information). This performance is manly limited by the detection speed of the LIA and the prediction fidelity of the neural network.

To improve the performance, the first way is to enhance the spin-state readout efficiency or use the ensemble NV centers. Then the detection speed is increased which enables sensing and reduction of higher-frequency noises such as the classical Overhauser field from $^{13}$C nuclear spins.The second way is to optimize the performance of prediction, \emph{i.e.}, trying more complicated DL models and providing more computational resources for training and processing. Comparing with coherent feedback scheme that often uses nuclear spins as ancillary qubits \cite{Cappellaro_feedback} or quantum memory \cite{Joerg_sensing_cluster}, our scheme requires no extra quantum resources and can work with general-purpose measurements. This approach can find applications in spin-based quantum computation, quantum simulation and nano-scale sensing (\emph{e.g.}, magnetic fields, temperature or electric fields), and be easily extended to other spin-based quantum systems, for example, the silicon-vacancy centers in diamond\cite{nc_siv_2017}, the defects in hexagonal boron nitride\cite{prl_hbn_2022} or the quantum dots\cite{nc_charge_2022}.

\begin{acknowledgement}

This work was supported by the National Key R\&D Program of China (Grant No. 2018YFA0306600), the National Natural Science Foundation of China (Grants No. 92265114 and 92265204), the CAS (Grants No. XDC07000000, GJJSTD20200001 and QYZDYSSW-SLH004), the Anhui Initiative in Quantum Information Technologies (Grant No. AHY050000), the Research Initiation Project (Grant No. K2022MB0PI02) and Center-initiated Research Project (Grant No. 2021MB0AL01) of Zhejiang Lab, and the Fundamental Research Funds for the Central Universities.

\end{acknowledgement}

\begin{suppinfo}

\begin{itemize}
  \item Supporting Information: Additional experimental datails, materials, and methods (PDF)
\end{itemize}

\end{suppinfo}

\providecommand{\latin}[1]{#1}
\makeatletter
\providecommand{\doi}
  {\begingroup\let\do\@makeother\dospecials
  \catcode`\{=1 \catcode`\}=2 \doi@aux}
\providecommand{\doi@aux}[1]{\endgroup\texttt{#1}}
\makeatother
\providecommand*\mcitethebibliography{\thebibliography}
\csname @ifundefined\endcsname{endmcitethebibliography}
  {\let\endmcitethebibliography\endthebibliography}{}

\end{document}